\documentclass[11pt]{article}
 \usepackage{graphicx}
\usepackage{tikz-cd}

\usepackage{graphics,graphicx,color,float, hyperref}
\usepackage{epsfig,bbm}
\usepackage{bm}
\usepackage{mathtools}
\usepackage{amsmath}
\usepackage{amssymb}
\usepackage{amsfonts}
\usepackage{amsthm}
\usepackage {times}
\usepackage{layout}
\usepackage{setspace}
\usepackage{rotating}

\newtheorem{fact}{Fact}

\newcommand{\beq}{\begin{equation}}
\newcommand{\enq}{\end{equation}}
\newcommand{\bel}{\begin{lemma}}
\newcommand{\enl}{\end{lemma}}
\newcommand{\bet}{\begin{theorem}}
\newcommand{\ent}{\end{theorem}}

\newcommand{\E}{\mathbb{E}}
\newcommand{\Tr}{\mathrm{Tr}}
\newcommand{\ketbra}[1]{|#1\rangle\langle#1|}

\newcommand{\eps}{\varepsilon}

\newcommand*{\cA}{\mathcal{A}}

\newcommand*{\cH}{\mathcal{H}}

\newcommand{\cP}{\mathcal{P}}

\newcommand{\supp}{\mathrm{supp}}
\newcommand{\suppress}[1]{}

\newcommand{\defeq}{\ensuremath{ := }}
\newcommand{\F}{\mathrm{F}}
\newcommand{\Pur}{\mathrm{P}}
\newcommand {\br} [1] {\ensuremath{ \left( #1 \right) }}

\newcommand {\minusspace} {\: \! \!}
\newcommand {\smallspace} {\: \!}
\newcommand {\fn} [2] {\ensuremath{ #1 \minusspace \br{ #2 } }}
\newcommand {\ball} [2] {\fn{\mathcal{B}^{#1}}{#2}}
\newcommand {\relent} [2] {\fn{\mathrm{D}}{#1 \middle\| #2}}
\newcommand {\dmax} [2] {\fn{\mathrm{D}_{\max}}{#1 \middle\| #2}}
\newcommand {\dmaxeps} [3] {\fn{\mathrm{D}^{#3}_{\max}}{#1 \middle\| #2}}
\newcommand {\mutinf} [2] {\fn{\mathrm{I}}{#1 \smallspace : \smallspace #2}}
\newcommand {\imax}{\ensuremath{\mathrm{I}_{\max}}}
\newcommand {\imaxeps}{\ensuremath{\mathrm{I}^{\varepsilon}_{\max}}}
\newcommand {\condmutinf} [3] {\mutinf{#1}{#2 \smallspace \middle\vert \smallspace #3}}
\newcommand {\hmin} [2] {\fn{\mathrm{H}_{\min}}{#1 \middle | #2}}
\newcommand {\hmineps} [3] {\fn{\mathrm{H}^{#3}_{\min}}{#1 \middle | #2}}
\newcommand {\hmax} [2] {\fn{\mathrm{H}_{\max}}{#1 \middle | #2}}
\newcommand {\hmaxeps} [3] {\fn{\mathrm{H}^{#3}_{\max}}{#1 \middle | #2}}
\newcommand {\dheps} [3] {\ensuremath{\mathrm{D}_{\mathrm{H}}^{#3}\left(#1 \| #2\right)}}
\newcommand {\dhalf} [2] {\ensuremath{\tilde{\mathrm{D}}_{\frac{1}{2}}\left(#1 \| #2\right)}}

\newcommand {\id} {\ensuremath{\mathrm{I}}}

\newcommand{\bra}[1]{\langle #1|}
\newcommand{\ket}[1]{|#1 \rangle}

\mathchardef\mhyphen="2D

\newcommand*{\renyi}{R\'{e}nyi }

\makeatletter
\newcommand*{\rom}[1]{\expandafter\@slowromancap\romannumeral #1@}
\makeatother

\mathchardef\mhyphen="2D

\newtheorem{definition}{Definition}
\newtheorem{claim}{Claim}
\newtheorem{cor}{Corollary}

\newtheorem{theorem}{Theorem}
\newtheorem{lemma}{Lemma}

\begin {document}

\title{A one-shot achievability result for quantum state redistribution}
\author{
Anurag Anshu\footnote{Centre for Quantum Technologies, National University of Singapore, Singapore. \texttt{a0109169@u.nus.edu}} \qquad
Rahul Jain\footnote{Centre for Quantum Technologies, National University of Singapore and MajuLab, UMI 3654, 
Singapore. \texttt{rahul@comp.nus.edu.sg}} \qquad 
Naqueeb Ahmad Warsi\footnote{Centre for Quantum Technologies, National University of Singapore and School of Physical and Mathematical Sciences, Nanyang Technological University, Singapore and IIITD, Delhi. \texttt{warsi.naqueeb@gmail.com}} 
}

\date{}
\maketitle

\begin{abstract} 

We study the problem of entanglement-assisted quantum state redistribution in the one-shot setting and provide a new achievability result on the quantum communication required. Our bounds are in terms of the max-relative entropy and the hypothesis testing relative entropy. We use the techniques of convex split and position-based decoding to arrive at our result. We show that our result is upper bounded by the result obtained in Berta, Christandl, Touchette (2016).
\end{abstract}

\section{Introduction}

Quantum communication finds its most natural expression in the \textit{coherent} framework, where a communication task should be achieved without affecting the correlation with the environment (which refers to the quantum systems not possessed by the communicating parties). A well known example of this is the task of \textit{quantum state merging}, introduced in the asymptotic i.i.d. setting by \cite{HorodeckiOW07}, which led to an operational understanding of the negativity of conditional quantum entropy and showed how entanglement was uniquely responsible for this phenomenon. This task was further studied in \cite{AbeyesingheDHW09}, leading to a protocol for distributed quantum source compression.

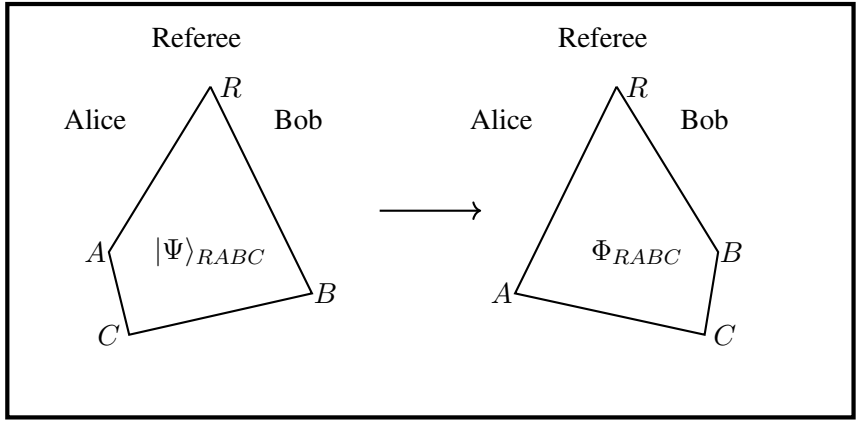
\begin{figure}[ht]
\centering
\begin{tikzpicture}[xscale=0.9,yscale=1.1]

\draw[ultra thick] (-3.5,6) rectangle (9,1);

\draw[thick] (-0.5,5) -- (-2,3) -- (-1.7,2) -- (1,2.5) -- (-0.5,5);
\node at (-0.2,5) {$R$};
\node at (-2.2,3) {$A$};
\node at (-2,2) {$C$};
\node at (1.2,2.5) {$B$};
\node at (-0.5,3) {$\ket{\Psi}_{RABC}$};

\node at (-0.7,5.6) {Referee};
\node at (-2.2,4.6) {Alice};
\node at (0.8,4.6) {Bob};

\draw[->, thick] (2,3.5) -- (3.5,3.5);

\draw[thick] (5.5,5) -- (4,2.5) -- (6.8,2) -- (7,3) -- (5.5,5);
\node at (5.8,5) {$R$};
\node at (3.8,2.5) {$A$};
\node at (7.2,3) {$B$};
\node at (7.1,2) {$C$};
\node at (5.8,3) {$\Phi_{RABC}$};

\node at (5.3,5.6) {Referee};
\node at (3.8,4.6) {Alice};
\node at (6.8,4.6) {Bob};

\end{tikzpicture}
\caption{\small The task of quantum state redistribution, where Alice needs to send her register $C$ to Bob, with the requirement that $\Pur(\Phi_{RABC},\ketbra{\Psi}_{RABC})\leq \eps$, for some error parameter $\eps$. Alice and Bob are allowed to have pre-shared entanglement. }
 \label{fig:stateredist}
\end{figure}

Quantum state merging serves as a special case of \textit{one-way} coherent quantum communication, where Alice (A), Bob (B) and Reference (R) share a joint pure quantum state $\ket{\Phi}_{RAB}$ and Alice needs to transmit her register $A$ to Bob, with the constraint that the Reference is not involved in the protocol and serves as the environment. But in a general communication scenario, Alice may not necessarily send all of the registers in her possession, suggesting a generalization of quantum state merging. This scenario is captured by the notion of \textit{quantum state redistribution} (Figure \ref{fig:stateredist}), first studied by~\cite{Devatakyard, YardD09} in the asymptotic i.i.d. setting. In this task, Alice possesses an additional register $C$ along with $A$ and needs to transfer $C$ to Bob.

Interestingly, quantum state merging and quantum state redistribution are closely related notions. The works \cite{YeBW08, oppenheim08} showed how to obtain a protocol for quantum state redistribution using a protocol for quantum state merging (and its \textit{time-reversed} version known as quantum state splitting). This relation also extends to the framework of \textit{one-shot} quantum information theory. The works \cite{Berta09} and \cite{Renner11} studied quantum state merging in the one-shot framework.  Using the aforementioned connection between quantum state merging and quantum state redistribution, the works ~\cite{DattaHO16, Berta14} obtained a one-shot bound on entanglement-assisted quantum communication cost of quantum state redistribution. These bounds were used by \cite{Dave14} to formulate a notion of \textit{quantum information complexity} and obtain a direct sum theorem for bounded round quantum communication complexity.

{\bf Our results:} In this work we provide a new achievability bound on the quantum communication cost of quantum state redistribution using entanglement-assisted one-shot protocols. Our bound (presented in Theorem \ref{newcompression}) is in terms of the max-relative entropy and the hypothesis testing relative entropy. This is in contrast to the achievability bound obtained in \cite{Berta14}, which is in terms of conditional max and min entropies. We also find in Theorem \ref{thm:compare} that our achievability bound is upper bounded by the corresponding bound in \cite{Berta14}. 

It was shown in \cite{Dave14} that the achievability result in \cite{Berta14} is upper bounded by $\frac{1}{2\eps^2}\condmutinf{R}{C}{B}$ (up to additive constants), where $\eps$ is an error parameter, $\condmutinf{R}{C}{B}$ is the quantum conditional mutual information and in present context is evaluated on the quantum state $\Phi_{RABC}$ on which quantum state redistribution has to be performed. Theorem \ref{thm:compare} thus allows us to conclude that our achievability result is upper bounded by $\frac{1}{2\eps^2}\condmutinf{R}{C}{B}_{\Phi}$ (up to multiplicative constants). The asymptotic behavior of our bound can be established by appealing to asymptotic equipartition properties of smooth max-relative entropy and hypothesis testing relative entropy \cite{TomHay13, li2014}, which we discuss in Theorem \ref{asymptoticstateredist}.

{\bf Techniques:} Our approach is different from those used in \cite{DattaHO16} and \cite{Berta14} (which are based on the technique of decoupling via a random unitary) and uses two ingredients. First is the technique of \textit{convex split} introduced in \cite{AnshuDJ14} in the context of compression of quantum messages (which also had implications for quantum state redistribution, further discussed in Section \ref{sec:compare}). This technique allows Alice to create a desired convex combination of quantum states on the registers of Bob and Reference. If Alice sent full information about this convex combination to Bob, he would simply output a correct register to finish the task. But this strategy would lead to a lot of communication from Alice. 

To circumvent this, we use the technique of \textit{quantum hypothesis testing}. This allows Alice to send limited information about the convex combination to Bob, after which he can gain the rest of the information by performing a quantum hypothesis testing on his registers. Details of the protocol appear in Section \ref{sec:quantumstate}, where Bob's decoding operation is a coherent version of the position-based decoding strategy introduced in \cite{AnshuJW17}. We note that recent works~\cite{AnshuJW17a,AnshuJW17b, AnshuJW17c,Wilde17a,QiWW17} have used similar techniques in various scenarios of quantum network theory. 

In Section \ref{sec:dhdhalf} we connect the hypothesis testing relative entropy to the sandwiched R\'{e}nyi relative entropy of order $\frac{1}{2}$. This allows our achievability result to be upper bounded by the difference between the max-relative entropy and the sandwiched R\'{e}nyi relative entropy of order $\frac{1}{2}$, which can further be upper bounded by the achievability result of \cite{Berta14} (Section \ref{sec:compare}). In order to connect  the hypothesis testing relative entropy to the sandwiched R\'{e}nyi relative entropy of order $\frac{1}{2}$, we consider the notion of \textit{pretty good measurement} introduced by Holevo \cite{Holevo73} (see also \cite{HJSWW96}) . We use the characterization given by Barnum and Knill \cite{BarnumK02} who showed the near optimality of this measurement. 

\subsection*{Organisation of the paper} 
We introduce our notations and notions used throughout the paper in Section \ref{sec:prelims}. In Section \ref{sec:quantumstate}, we present a new protocol for quantum state redistribution. In Section \ref{sec:compare}, we make a comparison of this protocol with previous works of \cite{Berta14} and \cite{AnshuDJ14}.
 
\section{Preliminaries}
\label{sec:prelims}

Consider a finite dimensional Hilbert space $\cH$ endowed with an inner product $\langle \cdot, \cdot \rangle$ (In this paper, we only consider finite dimensional Hilbert-spaces). The $\ell_1$ norm of an operator $X$ on $\cH$ is $\| X\|_1:=\Tr\sqrt{X^{\dagger}X}$ and $\ell_2$ norm is $\| X\|_2:=\sqrt{\Tr XX^{\dagger}}$. For hermitian operators $X, X'$, the notation $X\preceq X'$ implies that $X' - X$ is a positive semi-definite operator. A quantum state (or a density matrix or a state) is a positive semi-definite matrix on $\cH$ with trace equal to $1$. It is called {\em pure} if and only if its rank is $1$. A sub-normalized state is a positive semi-definite matrix on $\cH$ with trace less than or equal to $1$. Let $\ket{\psi}$ be a unit vector on $\cH$, that is $\langle \psi,\psi \rangle=1$.  With some abuse of notation, we use $\psi$ to represent the state and also the density matrix $\ketbra{\psi}$, associated with $\ket{\psi}$. Given a quantum state $\rho$ on $\cH$, the {\em support of $\rho$}, called $\text{supp}(\rho)$ is the subspace of $\cH$ spanned by all eigenvectors of $\rho$ with non-zero eigenvalues. For quantum states $\rho,\sigma$ on $\cH$, the notation $\text{supp}(\rho)\subseteq \text{supp}(\sigma)$ means that the support of $\rho$ is contained in the support of $\sigma$.

A {\em quantum register} $A$ is associated with some Hilbert space $\cH_A$. Define $|A| := \dim(\cH_A)$. Let $\mathcal{L}(A)$ represent the set of all linear operators acting on the set of quantum states on the Hilbert space $\cH_A$. We denote by $\mathcal{D}(A)$, the set of quantum states on the Hilbert space $\cH_A$. State $\rho$ with subscript $A$ indicates $\rho_A \in \mathcal{D}(A)$. If two registers $A,B$ are associated with the same Hilbert space, we shall represent the relation by $A\equiv B$.  Composition of two registers $A$ and $B$, denoted $AB$, is associated with Hilbert space $\cH_A \otimes \cH_B$.  For two quantum states $\rho\in \mathcal{D}(A)$ and $\sigma\in \mathcal{D}(B)$, $\rho\otimes\sigma \in \mathcal{D}(AB)$ represents the tensor product (Kronecker product) of $\rho$ and $\sigma$. The identity operator on $\cH_A$ (and associated register $A$) is denoted $\id_A$. 

Let $\rho_{AB} \in \mathcal{D}(AB)$. We define
\[ \rho_{B} := \Tr_{A}(\rho_{AB})
:= \sum_i (\bra{i} \otimes \id_{B})
\rho_{AB} (\ket{i} \otimes \id_{B}) , \]
where $\{\ket{i}\}_i$ is an orthonormal basis for the Hilbert space $\cH_A$.
The state $\rho_B\in \mathcal{D}(B)$ is referred to as the marginal state of $\rho_{AB}$. Unless otherwise stated, a missing register from subscript in a state will represent partial trace over that register. Given a $\rho_A\in\mathcal{D}(A)$, a {\em purification} of $\rho_A$ is a pure state $\rho_{AB}\in \mathcal{D}(AB)$ such that $\Tr_{B}(\rho_{AB})=\rho_A$. Purification of a quantum state is not unique.

A quantum {map} $\E: \mathcal{L}(A)\rightarrow \mathcal{L}(B)$ is a completely positive and trace preserving (CPTP) linear map (mapping states in $\mathcal{D}(A)$ to states in $\mathcal{D}(B)$). A {\em unitary} operator $U_A:\cH_A \rightarrow \cH_A$ is such that $U_A^{\dagger}U_A = U_A U_A^{\dagger} = \id_A$. An {\em isometry}  $V:\cH_A \rightarrow \cH_B$ is such that $V^{\dagger}V = \id_A$ and $VV^{\dagger} = \Pi_B$, where $\Pi_B$ is a projection on $\cH_B$. The set of all unitary operations on register $A$ is  denoted by $\mathcal{U}(A)$.

\begin{definition}
We shall consider the following information theoretic quantities. Reader is referred to ~\cite{Renner05, Tomamichel09,Tomamichel12,Datta09, GilchristLN05, WildeWY14, LennertDSFT13} for many of these definitions. We consider only normalized states in the definitions below. Let $\varepsilon \geq 0$. 
\begin{enumerate}
\item {\bf Fidelity} For $\rho_A,\sigma_A \in \mathcal{D}(A)$, $$\F(\rho_A,\sigma_A)\defeq\|\sqrt{\rho_A}\sqrt{\sigma_A}\|_1.$$ For classical probability distributions $P = \{p_i\}, Q =\{q_i\}$, $$\F(P,Q)\defeq \sum_i \sqrt{p_i \cdot q_i}.$$
\item {\bf Purified distance} For $\rho_A,\sigma_A \in \mathcal{D}(A)$, $$\Pur(\rho_A,\sigma_A) = \sqrt{1-\F^2(\rho_A,\sigma_A)}.$$
\item {\bf $\varepsilon$-ball} For $\rho_A\in \mathcal{D}(A)$, $$\ball{\eps}{\rho_A} \defeq \{\rho'_A\in \mathcal{D}(A)|~\Pur(\rho_A,\rho'_A) \leq \varepsilon\}. $$ 

\item {\bf Von-neumann entropy} For $\rho_A\in\mathcal{D}(A)$, $$S(\rho_A) \defeq - \Tr(\rho_A\log\rho_A) .$$ 
\item {\bf Relative entropy} For $\rho_A,\sigma_A\in \mathcal{D}(A)$ such that $\text{supp}(\rho_A) \subseteq \text{supp}(\sigma_A)$, $$\relent{\rho_A}{\sigma_A} \defeq \Tr(\rho_A\log\rho_A) - \Tr(\rho_A\log\sigma_A) .$$ 
\item {\bf Max-relative entropy} For $\rho_A,\sigma_A\in \mathcal{D}(A)$ such that $\text{supp}(\rho_A) \subseteq \text{supp}(\sigma_A)$, $$ \dmax{\rho_A}{\sigma_A}  \defeq  \inf \{ \lambda \in \mathbb{R} :   \rho_A \preceq 2^{\lambda} \sigma_A \}  .$$  
\item {\bf Quantum hypothesis testing relative entropy}  For $\rho_A,\sigma_A\in \mathcal{D}(A)$ and $\eps \in (0,1)$, $$ \mathrm{D}_{\mathrm{H}}^{\eps}(\rho_A\| \sigma_A)  \defeq  \sup_{0\preceq \Pi \preceq I, \Tr(\Pi\rho_A)\geq 1-\eps}\log\left(\frac{1}{\Tr(\Pi\sigma_A)}\right).$$  
\item {\bf Sandwiched Quantum \renyi relative entropy of order $\frac{1}{2}$} For $\rho_A,\sigma_A\in \mathcal{D}(A)$, $$ \dhalf{\rho_A}{\sigma_A}  \defeq   -2\log \F(\rho_A,\sigma_A) .$$
\item {\bf Mutual information} For $\rho_{AB}\in \mathcal{D}(AB)$, 
\begin{eqnarray*}
\mutinf{A}{B}_{\rho}&\defeq& S(\rho_A) + S(\rho_B)-S(\rho_{AB}) = \relent{\rho_{AB}}{\rho_A\otimes\rho_B}.
\end{eqnarray*}
\item {\bf Conditional mutual information} For $\rho_{ABC}\in\mathcal{D}(ABC)$, $$\condmutinf{A}{B}{C}_{\rho}\defeq \mutinf{A}{BC}_{\rho}-\mutinf{A}{C}_{\rho}.$$
\item {\bf Max-information}  For $\rho_{AB}\in \mathcal{D}(AB)$, $$ \imax(A:B)_{\rho} \defeq   \inf_{\sigma_{B}\in \mathcal{D}(B)}\dmax{\rho_{AB}}{\rho_{A}\otimes\sigma_{B}} .$$
\item {\bf Smooth max-information} For $\rho_{AB}\in \mathcal{D}(AB)$,  $$\imaxeps(A:B)_{\rho} \defeq \inf_{\rho'\in \ball{\eps}{\rho}} \imax(A:B)_{\rho'} .$$	
\item {\bf Conditional min-entropy} $$ \hmin{A}{B}_{\rho} \defeq  - \inf_{\sigma_B\in \mathcal{D}(B)}\dmax{\rho_{AB}}{\id_{A}\otimes\sigma_{B}} .$$  	
\item {\bf Conditional max-entropy} $$\hmax{A}{B}_{\rho} \defeq \max_{\sigma_B\in\mathcal{D}(B)}\log\F^2(\rho_{AB},\id_A\otimes \sigma_B).$$
\item {\bf Smooth conditional min-entropy} $$\hmineps{A}{B}{\eps}_{\rho} \defeq   \sup_{\rho^{'} \in \ball{\eps}{\rho}} \hmin{A}{B}_{\rho^{'}} .$$  	
\item {\bf Smooth conditional max-entropy} $$ \hmaxeps{A}{B}{\eps}_{\rho} \defeq \inf_{\rho^{'} \in \ball{\eps}{\rho}} \hmax{A}{B}_{\rho^{'}} .$$ 
\end{enumerate}
\label{def:infquant}
\end{definition}	

We will use the following facts. 
\begin{fact}[Triangle inequality for purified distance, ~\cite{GilchristLN05, Tomamichel12}]
\label{fact:trianglepurified}
For states $\rho_A, \sigma_A, \tau_A\in \mathcal{D}(A)$,
$$\Pur(\rho_A,\sigma_A) \leq \Pur(\rho_A,\tau_A)  + \Pur(\tau_A,\sigma_A) . $$ 
\end{fact}

\begin{fact}[\cite{stinespring55}](\textbf{Stinespring representation})\label{stinespring}
Let $\E(\cdot): \mathcal{L}(A)\rightarrow \mathcal{L}(B)$ be a quantum operation. There exists a register $C$ and an unitary $U\in \mathcal{U}(ABC)$ such that $\E(\omega)=\Tr_{A,C}\br{U (\omega  \otimes \ketbra{0}^{B,C}) U^{\dagger}}$. Stinespring representation for a channel is not unique. 
\end{fact}

\begin{fact}[Monotonicity under quantum operations, \cite{barnum96},\cite{lindblad75}]
	\label{fact:monotonequantumoperation}
For quantum states $\rho$, $\sigma \in \mathcal{D}(A)$, and quantum operation $\E(\cdot):\mathcal{L}(A)\rightarrow \mathcal{L}(B)$, it holds that
\begin{eqnarray*}
	 &&\F(\E(\rho),\E(\sigma)) \geq \F(\rho,\sigma) \quad \mbox{and} \quad \relent{\rho}{\sigma}\geq \relent{\E(\rho)}{\E(\sigma)}.
\end{eqnarray*}
In particular, for bipartite states $\rho_{AB},\sigma_{AB}\in \mathcal{D}(AB)$, it holds that
\begin{eqnarray*}
	&& \F(\rho_{AB},\sigma_{AB}) \leq \F(\rho_A,\sigma_A) \quad \mbox{and} \quad \relent{\rho_{AB}}{\sigma_{AB}}\geq \relent{\rho_A}{\sigma_A} .
\end{eqnarray*}
\end{fact}

\begin{fact}[Uhlmann's theorem, \cite{uhlmann76}]
\label{uhlmann}
Let $\rho_A,\sigma_A\in \mathcal{D}(A)$. Let $\rho_{AB}\in \mathcal{D}(AB)$ be a purification of $\rho_A$ and $\sigma_{AC}\in\mathcal{D}(AC)$ be a purification of $\sigma_A$. There exists an isometry $V: C \rightarrow B$ such that,
 $$\F(\ketbra{\theta}_{AB}, \ketbra{\rho}_{AB}) = \F(\rho_A,\sigma_A) ,$$
 where $\ket{\theta}_{AB} = (\id_A \otimes V) \ket{\sigma}_{AC}$.
\end{fact}

\begin{fact}[Gentle measurement lemma,\cite{Winter:1999,Ogawa:2002}]
\label{gentlelemma}
Let $\rho$ be a quantum state and $0 \preceq A \preceq I$ be an operator. Then 
$$\F(\rho, \frac{A\rho A}{\Tr(A^2\rho)})\geq \sqrt{\Tr(A^2\rho)}.$$
\end{fact}
\begin{proof}
Let $\ket{\rho}$ be a purification of $\rho$. Then $(I \otimes A)\ket{\rho}$ is a purification of $A\rho A$. Now, applying monotonicity of fidelity under quantum operations (Fact \ref{fact:monotonequantumoperation}), we find 
$$\F(\rho, \frac{A\rho A}{\Tr(A^2\rho)}) \geq \F(\ketbra{\rho}, \frac{(I \otimes A)\ketbra{\rho}(I\otimes A^{\dagger})}{\Tr(A^2\rho)}) = \sqrt{\frac{\Tr(A\rho)^2}{\Tr(A^2\rho)}} \geq \sqrt{\Tr(A^2\rho)}.$$
In last inequality, we have used $A^2 \preceq A$.
\end{proof}

\begin{fact}[Pretty-good measurement,\cite{BarnumK02}]
\label{prettygoodmeas}
Consider an ensemble $\{p_k,\rho^k_A\}$ such that $\sum_kp_k = 1$. Define $\rho_A=\sum_kp_k\rho^k$. Then it holds that 
$$\sum_k p^2_k\Tr(\rho_A^{-\frac{1}{2}}\rho^k_A\rho_A^{-\frac{1}{2}}\rho^k_A) \geq 1- \sum_{k\neq k'} \sqrt{p_kp_{k'}}\F(\rho^k_A,\rho^{k'}_A).$$

\end{fact}

\begin{fact}[Convex-split lemma, \cite{AnshuDJ14}]
\label{convexcomb}
Let $\rho_{PQ}\in\mathcal{D}(PQ)$ and $\sigma_Q\in\mathcal{D}(Q)$ be quantum states such that $\text{supp}(\rho_Q)\subseteq\text{supp}(\sigma_Q)$.  Let $k \defeq \dmax{\rho_{PQ}}{\rho_P\otimes\sigma_Q}$. Define the following state
\begin{eqnarray*}
\tau_{PQ_1Q_2\ldots Q_n} \defeq && \frac{1}{n}\sum_{j=1}^n \rho_{PQ_j}\otimes\sigma_{Q_1}\otimes \sigma_{Q_2}\ldots\otimes\sigma_{Q_{j-1}}\otimes\sigma_{Q_{j+1}}\ldots\otimes\sigma_{Q_n}
\end{eqnarray*}
on $n+1$ registers $P,Q_1,Q_2,\ldots Q_n$, where $\forall j \in [n]: \rho_{PQ_j} = \rho_{PQ}$ and $\sigma_{Q_j}=\sigma_Q$.  Then, for $\delta \in (0,1)$  
 $$ \F^{2}(\tau_{PQ_1Q_2\ldots Q_n},\tau_P \otimes \sigma_{Q_1}\otimes\sigma_{Q_2}\ldots \otimes \sigma_{Q_n}) \geq 1 - \delta,$$ 
 if $ n= \lceil\frac{2^k}{\delta}\rceil$.
\end{fact}

\begin{fact}[Hayashi-Nagaoka inequality, \cite{hayashinagaoka} ]
\label{haynag}
Let $0\preceq S\preceq \id$  and $T$ be positive semi-definite operators. Then 
$$\id - (S+T)^{-\frac{1}{2}}S(S+T)^{-\frac{1}{2}}\preceq 2(\id-S) + 4T.$$

\end{fact}

\suppress{
\begin{fact} [\cite{AnshuJW17}]
\label{closestatesmeasurement}
Let $\rho,\sigma$ be quantum states such that $\Pur(\rho,\sigma)\leq \eps$. Let $0\preceq \Pi\preceq \id$ be an operator such that $\Tr(\Pi\rho)\geq 1-\delta^2$. Then $\Tr(\Pi\sigma)\geq 1- (2\eps+\delta)^2$. If $\delta=0$, then $\Tr(\Pi\sigma) \geq 1-\eps^2$.
\end{fact}
}
The following fact was shown implicitly in \cite{Renner13} and used explicitly in \cite[Claim 5, Appendix A]{AnshuJW17} .
\begin{fact}[\cite{Renner13}]
\label{dmaxequivalence}
Let $\eps\in (0,1)$. For quantum states $\psi_{AB},\sigma_A,\sigma_B$, there exists a state $\bar{\psi}_{AB}\in\ball{\eps}{\psi_{AB}}$ such that 
$$\dmax{\bar{\psi}_{AB}}{\bar{\psi}_A\otimes \sigma_B} \leq \dmax{\psi_{AB}}{\sigma_A\otimes \sigma_B}+ \log\frac{3}{\eps^2}.$$
\end{fact}

\section{An achievability bound on quantum state redistribution}
\label{sec:quantumstate}

Quantum state redistribution is the following coherent quantum task (see Figure \ref{fig:stateredist}).

\textbf{Quantum state redistribution task}: Alice, Bob and Reference share a pure state $\ket{\Phi}_{RABC}$, with $AC$ belonging to Alice, $B$ to Bob and $R$ to Reference. Alice needs to transfer the register $C$ to Bob, such that the final state $\Phi'_{RABC}$ satisfies $\Pur(\Phi'_{RABC},\Phi_{RABC})\leq \eps$, for a given $\eps \in (0,1)$ which is the error parameter.  Alice and Bob are allowed to have pre-shared entanglement.

Following is the main result of this section. Observe the symmetry under the change of registers $B$ and $A$, which reflects the same property of conditional quantum mutual information first clarified in its operational interpretation by Devatak and Yard \cite{Devatakyard}, that is $\condmutinf{R}{C}{B}=\condmutinf{R}{C}{A}$.  
\begin{theorem}[Achievability bound]
\label{newcompression}
Fix $\eps_1,\eps_2\in (0,1)$ satisfying $3\eps_1+6\eps_2\leq 1$. There exists an entanglement-assisted one-way protocol $\cP$, which takes as input $\ket{\Phi}_{RACB}$ shared between three parties Reference ($R$), Bob ($B$) and Alice ($AC$) and outputs a state $\Phi'_{RACB}$ shared between Reference ($R$), Bob ($BC$) and Alice ($A$) such that $\Phi'_{RACB} \in  \ball{3\eps_1+6\eps_2}{\Phi_{RACB}}$ and the number of qubits communicated by Alice to Bob in $\cP$ is upper bounded by the minimum of the following quantities:
\begin{eqnarray*}
&&  \inf_{\sigma_C}\frac{1}{2}\bigg( \inf_{\Phi'_{RBC}\in \ball{\eps_1}{\Phi_{RBC}}}\dmax{\Phi'_{RBC}}{\Phi'_{RB}\otimes \sigma_C} - \sup_{\Phi''_{BC}\in \ball{\eps_2}{\Phi_{BC}}}\dheps{\Phi''_{BC}}{\Phi''_{B}\otimes \sigma_C}{\eps^2_2}\bigg)  \\ && +  \log\left(\frac{1}{\varepsilon_1\cdot \eps_2}\right)
\end{eqnarray*}
and 
\begin{eqnarray*}
&& \inf_{\sigma_C}\frac{1}{2}\bigg( \inf_{\Phi'_{RAC}\in \ball{\eps_1}{\Phi_{RAC}}}\dmax{\Phi'_{RAC}}{\Phi'_{RA}\otimes \sigma_C} - \sup_{\Phi''_{AC}\in \ball{\eps_2}{\Phi_{AC}}}\dheps{\Phi''_{AC}}{\Phi''_{A}\otimes \sigma_C}{\eps^2_2}\bigg)\\ &&+ \log\left(\frac{1}{\varepsilon_1\cdot \eps_2}\right).
\end{eqnarray*}
\end{theorem}

\vspace{0.1in}

\noindent\textit{Outline of the proof:} The proof of Theorem \ref{newcompression} is obtained by combining the convex-split technique from \cite{AnshuDJ14} and position-based decoding technique from \cite{AnshuJW17a}. Alice, Bob and Reference share the quantum state $\ket{\Phi}_{RABC}$. Furthermore, Alice and Bob share $n$ copies of a purification of the quantum state $\sigma_C$, where $\log n \approx \dmaxeps{\Phi_{RBC}}{\Phi_{RB}\otimes \sigma_C}{\eps_1}$ (the global state is $\ket{\xi}$ in Equation \ref{eq:statesinprot}). By performing an appropriate measurement on her registers, Alice is able to prepare a quantum state close to $\mu$ (defined in Equation \ref{eq:statesinprot}) on the registers of Bob and Reference. This is possible due to the convex-split lemma (Lemma \ref{convexcomb}) and the choice of $n$. Moreover, the index $j$ appearing in the definition of $\mu$ is her measurement outcome. If she could communicate $j$ to Bob, he would be able to pick up the register $C_j$ obtaining the quantum state $\Phi_{RBC_j}$. Since $\Phi_{RBC_j}$ is independent of the quantum state in registers $C_1,\ldots C_{j-1}, C_{j+1}, \ldots C_n$ (conditioned on measurement outcome $j$), its purification lies on Alice's registers. This would allow Alice to apply appropriate isometry, obtaining the desired quantum state $\ket{\Phi}_{RABC_j}$.  

\begin{figure}[h]
\centering
\begin{tikzpicture}[xscale=0.9,yscale=1.2]

\node at (2.5,5) {$\frac{1}{b}$};

\draw[fill] (3.5,7) circle [radius=0.05];
\draw[fill] (3.5,8) circle [radius=0.05];
\draw[thick] (3.5,8) to [out=200,in=160] (3.5,7);
\node at (2.4,7.5) {$\Psi_{RBAC_1}$};
\node at (3.5,6.5) {$\otimes$};
\node at (2.9,6) {$\sigma_{C_2}$};
\draw[fill] (3.5,6) circle [radius=0.05];
\node at (3.5,5.5) {$\otimes$};
\draw[fill] (3.5,5) circle [radius=0.05];
\node at (3.5,4.5) {$\otimes$};
\draw[fill] (3.5,4.1) circle [radius=0.02];
\draw[fill] (3.5,3.7) circle [radius=0.02];
\draw[fill] (3.5,3.3) circle [radius=0.02];
\draw[fill] (3.5,2.9) circle [radius=0.02];
\node at (3.5,2.5) {$\otimes$};
\draw[fill] (3.5,2) circle [radius=0.05];
\node at (2.9,2) {$\sigma_{C_b}$};

\node at (4.5,5) {$+$};

\node at (5.5,5) {$\frac{1}{b}$};

\draw[fill] (6.5,7) circle [radius=0.05];
\draw[fill] (6.5,8) circle [radius=0.05];
\node at (7.1,7) {$\sigma_{C_1}$};
\draw[thick] (6.5,8) to [out=200,in=160] (6.5,6);
\node at (5.1,7) {$\Psi_{RBAC_2}$};
\node at (6.5,6.5) {$\otimes$};
\draw[fill] (6.5,6) circle [radius=0.05];
\node at (6.5,5.5) {$\otimes$};
\draw[fill] (6.5,5) circle [radius=0.05];
\node at (6.5,4.5) {$\otimes$};
\draw[fill] (6.5,4.1) circle [radius=0.02];
\draw[fill] (6.5,3.7) circle [radius=0.02];
\draw[fill] (6.5,3.3) circle [radius=0.02];
\draw[fill] (6.5,2.9) circle [radius=0.02];
\node at (6.5,2.5) {$\otimes$};
\draw[fill] (6.5,2) circle [radius=0.05];
\node at (5.9,2) {$\sigma_{C_b}$};

\draw[fill] (7.5,5.0) circle [radius=0.02];
\draw[fill] (7.9,5.0) circle [radius=0.02];
\draw[fill] (8.3,5.0) circle [radius=0.02];
\draw[fill] (8.7,5.0) circle [radius=0.02];

\node at (9.2,5) {$+$};

\node at (9.7,5) {$\frac{1}{b}$};

\draw[fill] (10.5,7) circle [radius=0.05];
\draw[fill] (10.5,8) circle [radius=0.05];
\draw[thick] (10.5,8) to [out=230,in=85] (10.1,7) to [out=265,in=95] (10.1,3) to [out=275,in=130] (10.5,2);
\node at (9.4,7.5) {$\Psi_{RBAC_b}$};
\node at (11.1,7) {$\sigma_{C_1}$};
\node at (10.5,6.5) {$\otimes$};
\node at (11.1,6) {$\sigma_{C_2}$};
\draw[fill] (10.5,6) circle [radius=0.05];
\node at (10.5,5.5) {$\otimes$};
\draw[fill] (10.5,5) circle [radius=0.05];
\node at (10.5,4.5) {$\otimes$};
\draw[fill] (10.5,4.1) circle [radius=0.02];
\draw[fill] (10.5,3.7) circle [radius=0.02];
\draw[fill] (10.5,3.3) circle [radius=0.02];
\draw[fill] (10.5,2.9) circle [radius=0.02];
\node at (10.5,2.5) {$\otimes$};
\draw[fill] (10.5,2) circle [radius=0.05];

\end{tikzpicture}
\caption{\small Bob performs quantum hypothesis testing on the state $\mu^{(2)}_{RABC_1C_2\ldots C_b}$ (depicted above). This is the state he obtains after receiving Alice's message.}
 \label{fig:convexcomb}
\end{figure}
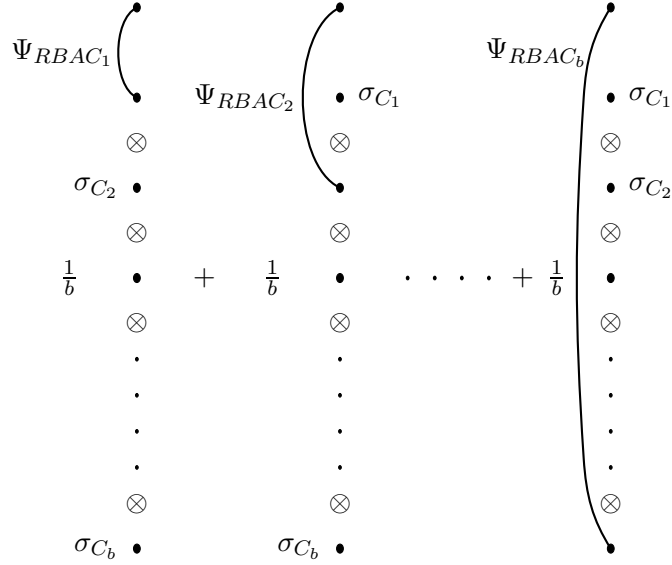

The problem is that the number of qubits required to communicate $j$ is large ($\approx \frac{1}{2}\dmaxeps{\Phi_{RBC}}{\Phi_{RB}\otimes \sigma_C}{\eps_1}$). To circumvent this, Bob makes use of his quantum side information (that is the register $B$). Instead of communicating the value of $j$ to Bob, Alice only sends the value $\lfloor j/ b \rfloor$ to Bob, where $\log b \approx \dheps{\Phi_{BC}}{\Phi_B\otimes \sigma_C}{\eps_2}$. This reduces the communication to $\log n/b$, which is $$\approx \frac{1}{2}\left(\dmaxeps{\Phi_{RBC}}{\Phi_{RB}\otimes \sigma_C}{\eps_1} - \dheps{\Phi_{BC}}{\Phi_B\otimes \sigma_C}{\eps_2}\right).$$ Bob's task is to recover the actual value of $j$ given this limited information. Observe that upon receiving the message from Alice, the quantum state on the registers of Bob is close to the quantum state depicted in Figure \ref{fig:convexcomb}. Bob uses position-based decoding strategy to find the value of $j$, which is possible due to the chosen value of $b$. We take some additional care to make our protocol coherent. This allows us to consider a similar protocol where Bob sends register $C$ to Alice, and reverse it to achieve the task in Theorem \ref{newcompression}. Taking a minimum over the two communication costs, we obtain our achievability bound.
 
\vspace{0.1in}

\begin{proof}[Proof of Theorem \ref{newcompression}]

The proof is divided into the following parts.

\vspace{0.1in}

\noindent {\bf 1. Quantum state and registers appearing in the proof:} Let $\sigma_C$ be an arbitrary state in $\mathcal{D}(C)$. Let $k \defeq \inf_{\Phi'_{RBC}\in\ball{\eps_1}{\Phi_{RBC}}}\dmax{\Phi'_{RBC}}{\Phi'_{RB}\otimes \sigma_C}$, $\delta\defeq\varepsilon_1^2$ and $n \defeq \lceil\frac{2^k }{\delta}\rceil$. Fix a $\Phi''_{BC}\in \ball{\eps_2}{\Phi_{BC}}$, let $$b\defeq \lceil \eps^2_2\cdot 2^{\dheps{\Phi''_{BC}}{\Phi''_{B}\otimes \sigma_C}{\eps^2_2}}\rceil$$ and let 
\begin{equation}
\label{eq:hypooperator}
\Pi_{BC} := \underset{\Pi}{\arg\max} \left(\dheps{\Phi''_{BC}}{\Phi''_{B}\otimes \sigma_C}{\eps^2_2}\right)
\end{equation}
be the operator obtaining the supremum in the definition of $\dheps{\Phi''_{BC}}{\Phi''_{B}\otimes \sigma_C}{\eps^2_2}$. Define the following quantum states, 
\begin{eqnarray}
\label{eq:statesinprot}
&&\mu_{RB C_1 \ldots C_n} \defeq \frac{1}{n}\sum_{j=1}^n \Phi_{RBC_j}\otimes\sigma_{C_1}\otimes\ldots\otimes\sigma_{C_{j-1}}\otimes\sigma_{C_{j+1}}\otimes\ldots \otimes\sigma_{C_n},\nonumber\\ &&
\xi_{RB C_1\ldots C_n}  \defeq \Phi_{RB}\otimes \sigma_{C_1}\ldots \otimes \sigma_{C_n},\nonumber\\ &&
\ket{\theta}_{L_1\ldots L_n C_1\ldots C_n} = \ket{\sigma}_{L_1C_1}\otimes \ket{\sigma}_{L_2C_2}\ldots \ket{\sigma}_{L_nC_n},\nonumber\\ &&
\ket{\xi}_{RABC L_1\ldots L_nC_1\ldots C_n} \defeq  \ket{\Phi}_{RABC} \otimes \ket{\theta}_{L_1\ldots L_n C_1\ldots C_n}. 
\end{eqnarray}
Above, $\ket{\sigma}_{LC}$ is a purification of $\sigma_C$ in a register $L$. Note that $\Phi_{RB} = \mu_{RB}$.  Using Claim ~\ref{convexcombcor} (variant of convex split lemma) and choice of $n$ we have,
$$\F^2(\xi_{RBC_1\ldots C_n}, \mu_{RB C_1\ldots C_n})  \geq 1 - 9\varepsilon^2_1.$$ 
Consider the following purification of $\mu_{RB C_1\ldots C_n}$, 
\begin{eqnarray*}
\ket{\mu}_{RBAJL_1\ldots L_n C_1\ldots C_n} &=& \frac{1}{\sqrt{n}}\sum_{j=1}^n\ket{j}_J\ket{\Phi}_{RBAC_j}\otimes \ket{\sigma}_{L_1C_1}\otimes \ldots\otimes\ket{\sigma}_{L_{j-1}C_{j-1}}\otimes\ket{0}_{L_j}\otimes\\ &&\ket{\sigma}_{L_{j+1}C_{j+1}}\otimes\ldots\otimes\ket{\sigma}_{L_nC_n}. 
\end{eqnarray*} 
Let $\ket{\xi'}_{RBAJ L_1\ldots L_nC_1\ldots C_n}$ be a purification of $\xi_{RB C_1\ldots C_n}$ (guaranteed by Uhlmann's theorem, Fact~\ref{uhlmann}) such that,
\begin{eqnarray}
\label{eq:convexpur}
&&\F^2(\xi'_{RBAJL_1\ldots L_nC_1\ldots C_n},\mu_{RBAJ L_1\ldots L_nC_1\ldots C_n}) \nonumber \\ &&=  \F^2(\xi_{RBC_1\ldots C_n}, \mu_{RB C_1\ldots C_n})  \geq 1- 9\varepsilon^2_1.
\end{eqnarray}
Let $V': ACL_1\ldots L_n \rightarrow JA L_1\ldots L_n$ be an isometry such that,
$$V'\ket{\xi}_{RABC L_1\ldots L_nC_1\ldots C_n} = \ket{\xi'}_{RBAJ L_1\ldots L_nC_1\ldots C_n} .$$

\noindent {\bf 2. Construction of the protocol:} Now we proceed to construct the protocol $\cP$ as follows.
\begin{enumerate}
\item Alice, Bob and Reference start by sharing the state $\ket{\xi}_{RABC L_1\ldots L_nC_1\ldots C_n} $ between themselves where Alice holds registers $ACL_1\ldots L_n$, Reference holds the register $R$ and Bob holds the registers $BC_1\ldots C_n$. Note that  $\ket{\Phi}_{RABC}$ is provided as input to the protocol and $\ket{\theta}_{L_1\ldots L_n C_1\ldots C_n}$ is additional shared entanglement between Alice and Bob. 
\item Alice applies the isometry $V'$ to obtain the state $\ket{\xi'}_{RBAJL_1\ldots L_nC_1\ldots C_n}$, where Alice holds the registers $JAL_1\ldots L_n$, Reference holds the register $R$ and Bob holds the registers  $BC_1\ldots C_n$.
\begin{itemize}
\item At this stage, the global quantum state is close to the quantum state $\ket{\mu}_{RBAJL_1\ldots L_n C_1\ldots C_n}$ due to Equation \ref{eq:convexpur}.
\end{itemize}
\item Alice introduces two registers $J_1,J_2$ with $|J_1|= \lfloor n/b \rfloor$ and $|J_2| = b$. She applies an isometry $W:\cH_{J}\rightarrow \cH_{J_1}\otimes \cH_{J_2}$ such that $$W\ket{j}_J = \ket{\lfloor (j-1)/b \rfloor}_{J_1}\ket{ j \hspace{1mm} \% \hspace{1mm} b }_{J_2},$$ where $ j \hspace{1mm} \% \hspace{1mm} b $ is equal to $j\text{ mod } b$ (if $j\text{ mod } b\neq 0$) and equal to $b$ otherwise. Here $j\text{ mod } b$ is the remainder obtained by dividing $j$ with $b$.

\item Alice introduces a register $J'_1 \equiv J_1$ in the state $\ket{1}_{J'_1}$ and performs the operation $$\ket{j_1}_{J_1}\ket{1}_{J'_1}\rightarrow \ket{j_1}_{J_1}\ket{j_1}_{J'_1}.$$ She sends register $J'_1$ to Bob using $\frac{1}{2}\log \lfloor n/b \rfloor \leq \frac{\log (n/b)}{2}$ qubits of quantum communication. Alice and Bob employ superdense coding (\cite{bennett92}) using fresh shared entanglement to achieve this. 

\item Controlled on the value $j_1$ in the register $J'_1$, Bob swaps the set of registers $C_{b\cdot j_1+1}$, $C_{b\cdot j_1+2}$, $\ldots C_{b\cdot j_1+b}$ with the set of registers $C_1,C_2,\ldots C_b$ in that order. Alice swaps the set of registers $L_{b\cdot j_1+1}$, $L_{b\cdot j_1+2}$, $\ldots L_{b\cdot j_1+b}$ with the set of registers $L_1,L_2,\ldots L_b$ in that order. 

\begin{itemize}
\item If the quantum state $\ket{\mu}_{RBAJL_1\ldots L_nC_1\ldots C_n}$ was shared between Alice, Bob and Reference at Step $2$, then the joint state at this step of the protocol in the registers $RBAJ_2L_1L_2\ldots L_bC_1C_2\ldots C_b$ would be equal to (see Figure \ref{fig:convexcomb} for the marginal of this quantum state on registers with Bob)
\begin{eqnarray}
\label{eq:mu2def}
\ket{\mu^{(2)}}_{RBAJ_2L_1L_2\ldots L_bC_1C_2\ldots C_b} &:=&  \frac{1}{\sqrt{b}}\sum_{j_2=1}^b \ket{j_2}_{J_2}\ket{\Phi}_{RBAC_{j_2}}\otimes \ket{\sigma}_{L_1C_1}\otimes\ldots \nonumber\\ &&\ket{\sigma}_{L_{j_2-1}C_{j_2-1}}\otimes \ket{\sigma}_{L_{j_2+1}C_{j_2+1}}\otimes\ldots \otimes\ket{\sigma}_{L_bC_b} \nonumber\\ &&
\end{eqnarray}
\end{itemize}

\item Define the position-based operators \cite{AnshuJW17a}:
\begin{eqnarray*}
\Pi_{1}&\defeq& \Pi_{BC_1}\otimes \id_{C_2}\otimes\ldots \otimes \id_{C_b},\\
\Pi_{2}&\defeq& \id_{C_1}\otimes \Pi_{BC_2}\otimes \id_{C_3}\otimes\ldots \otimes \id_{C_b},\\
&&\vdots \\
\Pi_b &\defeq& \id_{C_1}\otimes \id_{C_2}\otimes\ldots \otimes \id_{C_{b-1}}\otimes \Pi_{BC_b},
\end{eqnarray*} 
where $\Pi_{BC}$ has been defined in Equation \ref{eq:hypooperator}. Let $\Pi\defeq \sum_{j_2} \Pi_{j_2}$. Bob performs the following isometry: $$V_B : = \sum_{j_2}\sqrt{\Pi^{-\frac{1}{2}}\Pi_{j_2}\Pi^{-\frac{1}{2}}} \otimes \ket{j_2}_{J'_2} + \sqrt{\id - \Pi^0}\ket{0}_{J'_2},$$ where $\Pi^0$ is the projector onto the support of $\Pi$ and $\ket{0}$ represents the possibility that no output in the set $\{1,2,\ldots b\}$ may be obtained. Then he swaps registers $C_{j_2}$ and $C_1$, controlled on values $\{1,2,\ldots b\}$ on the register $J'_2$ and does nothing for the value $0$.
\item Final state is obtained in the registers $RABC_1$. 

\end{enumerate}

\vspace{0.1in}

\noindent{\bf 3. Analysis of the protocol:} Let $\Phi'_{RABC_1}$ be the final quantum state in registers $RABC_1$. Let $\Phi^1_{RABC_1}$ be the quantum state obtained in registers $RABC_1$ if the quantum state $\ket{\mu}_{RBAJL_1\ldots L_nC_1\ldots C_n}$ was shared between Alice, Bob and Reference at Step $2$ of the protocol $\cP$. We now show that $\Pur(\Phi'_{RABC_1}, \Phi_{RABC_1})\leq 3\eps_1+6\eps_2$. Towards this, consider
\begin{eqnarray*}
\Pur(\Phi'_{RABC_1}, \Phi_{RABC_1})&&\overset{(1)}\leq\Pur(\Phi'_{RABC_1}, \Phi^1_{RABC_1}) + \Pur(\Phi_{RABC_1}, \Phi^1_{RABC_1}) \\
&&\overset{(2)}\leq 3\eps_1+ \Pur(\Phi_{RABC_1}, \Phi^1_{RABC_1}) \\
&& \overset{(3)}\leq 3\eps_1+6\eps_2,
\end{eqnarray*}
where $(1)$ follows from triangle inequality for purified distance (Fact \ref{fact:trianglepurified}); $(2)$ follows by applying the monotonicity of fidelity under quantum operation (Fact~\ref{fact:monotonequantumoperation})) in Equation \ref{eq:convexpur} to obtain
\begin{eqnarray*}
\F^2(\Phi^1_{RABC}, \Phi'_{RABC}) & \geq& \F^2(\xi'_{RBAJ L_1\ldots L_nC_1\ldots C_n},\mu_{RBAJ L_1\ldots L_nC_1\ldots C_n}) \nonumber\\ &\geq& 1- 9\varepsilon^2_1;
\end{eqnarray*}
and $(3)$ follows from the following Claim, which is proved towards the end.
\begin{claim}
\label{claimprotp1}
 It holds that $\Pur(\Phi^1_{RBAC_1}, \Phi_{RBAC_1}) \leq 6\eps_2$.
\end{claim}

Thus, we have shown that $\Phi'_{RABC_1}\in \ball{3\eps_1+6\eps_2}{\Phi_{RABC_1}}$. Furthermore, the number of qubits communicated by Alice to Bob in $\cP$ is equal to $\frac{\log(n/b)}{2}$. This is upper bounded by:
\begin{eqnarray*}
&&\frac{1}{2}\bigg(\inf_{\Phi'_{RBC}\in\ball{\eps_1}{\Phi_{RBC}}}\dmax{\Phi'_{RBC}}{\Phi'_{RB}\otimes \sigma_C} -   \sup_{\Phi''_{BC}\in \ball{\eps_2}{\Phi_{BC}}}\dheps{\Phi''_{BC}}{\Phi''_{B}\otimes \sigma_C}{\eps_2}\bigg)\\ &&+ \log\left(\frac{1}{\varepsilon_1\cdot \eps_2}\right) .
\end{eqnarray*}

A similar protocol $\cP'$ can be obtained where the register $C$ is originally with Bob and Bob sends his register $C$ to Alice. Since all the operations by Alice and Bob are isometries in the protocol $\cP'$, by reversing it one can achieve the task as stated in the theorem. This gives us the following upper bound on the number of qubits communicated: 

\begin{eqnarray*}
&&\frac{1}{2}\bigg( \inf_{\Phi'_{RAC}\in \ball{\eps_1}{\Phi_{RAC}}}\dmax{\Phi'_{RAC}}{\Phi'_{RA}\otimes \sigma_C}  - \sup_{\Phi''_{AC}\in \ball{\eps_2}{\Phi_{AC}}}\dheps{\Phi''_{AC}}{\Phi''_{A}\otimes \sigma_C}{\eps_2} \bigg) \\ &&+ \log\left(\frac{1}{\varepsilon_1\cdot\eps_2}\right).
\end{eqnarray*}

This gives the desired upper bound on the number of qubits communicated. To complete the proof of the theorem, we now establish the proof of Claim \ref{claimprotp1}.

\noindent {\bf Proof of Claim \ref{claimprotp1}:} Let $\ket{\Phi''}_{RABC}$ be a purification of $\Phi''_{BC}$ such that $\Pur(\ketbra{\Phi''}_{RABC}, \ketbra{\Phi}_{RABC}) = \Pur(\Phi''_{BC}, \Phi_{BC})$, as guaranteed by Uhlmann's Theorem (Fact \ref{uhlmann}). We consider the action of Bob's operation $V_B$ and the subsequent swap operation on the quantum state 
\begin{eqnarray*}
\ket{\mu''}_{RBAJ_2L_1L_2\ldots L_bC_1C_2\ldots C_b} &:=& \frac{1}{\sqrt{b}}\sum_{j_2=1}^b \ket{j_2}_{J_2}\ket{\Phi''}_{RBAC_{j_2}}\otimes \ket{\sigma}_{L_1C_1}\otimes\ldots \\ &&\ket{\sigma}_{L_{j_2-1}C_{j_2-1}}\otimes \ket{\sigma}_{L_{j_2+1}C_{j_2+1}}\otimes\ldots \otimes\ket{\sigma}_{L_bC_b}.
\end{eqnarray*}
 Since $\Phi''_{BC}\in\ball{\eps_2}{\Phi_{BC}}$, we have that $\Pur(\mu'',\mu^{(2)})\leq \eps_2$, where the quantum state $\ket{\mu^{(2)}}$ has been defined in Equation \ref{eq:mu2def}. Define the quantum state

\begin{eqnarray*}
\ket{\mu''_f}_{RBAJ_2J'_2L_1L_2\ldots L_bC_1C_2\ldots C_b} &:=& \frac{1}{\sqrt{b}}\sum_{j_2=1}^b \ket{j_2}_{J_2}\ket{j_2}_{J'_2}\ket{\Phi''}_{RBAC_{j_2}}\otimes \ket{\sigma}_{L_1C_1}\otimes\ldots \\ &&\ket{\sigma}_{L_{j_2-1}C_{j_2-1}}\otimes \ket{\sigma}_{L_{j_2+1}C_{j_2+1}}\otimes\ldots \otimes\ket{\sigma}_{L_bC_b}.
\end{eqnarray*}

We first prove that $\Pur(V_B \mu'' V_B^{\dagger}, \mu''_f) \leq 4\eps_2$ . Define the conditional probability distribution 
\begin{eqnarray*}
p_{j'_2|j_2} &\defeq& \Tr\bigg(\Pi^{-\frac{1}{2}}\Pi_{j'_2} \Pi^{-\frac{1}{2}} \Phi''_{BAC_{j_2}}\otimes \sigma_{C_1}\otimes\ldots \sigma_{C_{j_2-1}}\otimes \sigma_{C_{j_2+1}} \otimes \ldots \otimes\sigma_{C_b}\bigg),
\end{eqnarray*}
\begin{eqnarray*}
p_{0|j_2} &\defeq& \Tr\bigg((\id - \Pi^0) \Phi''_{BAC_{j_2}}\otimes \sigma_{C_1}\otimes\ldots \sigma_{C_{j_2-1}}\otimes \sigma_{C_{j_2+1}} \otimes \ldots \otimes\sigma_{C_b}\bigg).
\end{eqnarray*}
 From Claim \ref{gentlepovm}, we have 
\begin{eqnarray*}
\F(V_B \mu'' V_B^{\dagger}, \mu''_f) \geq \frac{1}{b}\sum_{j_2} p_{j_2|j_2} = 1- \frac{1}{b}\sum_{j'_2\neq j_2} p_{j'_2|j_2} = 1- \sum_{j'_2\neq 1} p_{j'_2|1}, 
\end{eqnarray*} 
where the last equality follows from symmetry under change of $j_2$. Then 
\begin{eqnarray*}
\sum_{j'_2\neq 1} p_{j'_2|1}  &=& \Tr\bigg((\id -\Pi^{-\frac{1}{2}}\Pi_{1} \Pi^{-\frac{1}{2}}) \Phi''_{BAC_1}\otimes \sigma_{C_2}\otimes \ldots \otimes\sigma_{C_b}\bigg)\\  &\overset{(1)}\leq& 2\cdot\Tr\left((\id - \Pi_{1}) \Phi''_{BAC_1}\otimes \sigma_{C_2}\otimes \ldots \otimes\sigma_{C_b}\right) \\ &+& 4\cdot \sum_{j_2\neq 1} \Tr\left(\Pi_{j_2} \Phi''_{BAC_1}\otimes \sigma_{C_2}\otimes \ldots \otimes\sigma_{C_b}\right)\\ &\overset{(2)}\leq& 2\cdot \eps^2_2 + 4\cdot b\cdot 2^{- \dheps{\Phi''_{BC}}{\Phi''_B\otimes \sigma_C}{\eps^2_2}} \\ &\overset{(3)}\leq& 6\eps_2^2,
\end{eqnarray*}
where $(1)$ follows from the Hayashi-Nagaoka inequality (Fact \ref{haynag}), $(2)$ follows from Equation \ref{eq:hypooperator} and $(3)$ follows from the choice of $b$. This implies that $\F(V_B \mu'' V_B^{\dagger}, \mu''_f) \geq 1- 6\eps_2^2$, from which we conclude the desired relation $\Pur(V_B \mu'' V_B^{\dagger}, \mu''_f) \leq 4\eps_2$. 

If Bob swaps the registers $C_j$ and $C_1$ in the quantum state $\mu''_f$ (controlled on the value in register $J'$), the output in registers $RABC_1$ is $\ket{\Phi''}_{RABC_1}$. Since $\Phi''_{RABC_1}\in \ball{\eps_2}{\Phi_{RABC_1}}$,  $\Pur(V_B \mu'' V_B^{\dagger}, \mu''_f) \leq 4\eps_2$ and $\Pur(\mu'',\mu^{(2)}) \leq \eps_2$, we conclude using monotonicity of purified distance under quantum operations (Fact \ref{fact:monotonequantumoperation}) and triangle inequality for purified distance (Fact \ref{fact:trianglepurified}) that $\Pur(\Phi^1_{RABC_1}, \Phi_{RABC_1})\leq 6\eps_2$.

\vspace{0.1in}

\end{proof}

\subsection{Claims used in the proof of Theorem \ref{newcompression}}

\begin{claim}[A variant of convex split lemma]
\label{convexcombcor}
Fix an $\eps\in (0,1)$. Let $\rho_{PQ}\in\mathcal{D}(PQ)$ and $\sigma_Q\in\mathcal{D}(Q)$ be quantum states such that $\text{supp}(\rho_Q)\subseteq\text{supp}(\sigma_Q)$.  Let $k \defeq \inf_{\rho'_{PQ}\in\ball{\eps}{\rho_{PQ}}}\dmax{\rho'_{PQ}}{\rho'_P\otimes\sigma_Q}$. Define the following state
\begin{eqnarray*}
\tau_{PQ_1Q_2\ldots Q_n} \defeq  &&\frac{1}{n}\sum_{j=1}^n \rho_{PQ_j}\otimes\sigma_{Q_1}\otimes \sigma_{Q_2}\ldots\otimes\sigma_{Q_{j-1}} \otimes\sigma_{Q_{j+1}}\ldots\otimes\sigma_{Q_n}
\end{eqnarray*}
on $n+1$ registers $P,Q_1,Q_2,\ldots Q_n$, where $\forall j \in [n]: \rho_{PQ_j} = \rho_{PQ}$ and $\sigma_{Q_j}=\sigma_Q$.  For $\delta \in (0,(1-2\eps)^2) $ and $ n= \lceil\frac{2^k}{\delta}\rceil$, it holds that 
\begin{eqnarray*}
&&\F^2(\tau_{PQ_1Q_2\ldots Q_n},\tau_P \otimes \sigma_{Q_1}\otimes\sigma_{Q_2}\ldots \otimes \sigma_{Q_n}) \geq 1-  (\sqrt{\delta}+2\eps)^2.
\end{eqnarray*}
\end{claim}
\begin{proof}

Let $\rho'_{PQ}$ be the state achieving the infimum in $k$. Consider the state
\begin{eqnarray*}
\tau'_{PQ_1Q_2\ldots Q_n} \defeq && \frac{1}{n}\sum_{j=1}^n \rho'_{PQ_j}\otimes\sigma_{Q_1}\otimes \sigma_{Q_2}\ldots\otimes\sigma_{Q_{j-1}}\otimes\sigma_{Q_{j+1}}\ldots\otimes\sigma_{Q_n}
\end{eqnarray*}

From Fact \ref{convexcomb} we have that $\F^2(\tau'_{PQ_1Q_2\ldots Q_n}, \rho'_P\otimes\sigma_{Q_1}\otimes\ldots \sigma_{Q_n})\geq 1-\delta$. Now, $\F^2(\tau'_{PQ_1Q_2\ldots Q_n},\tau_{PQ_1Q_2\ldots Q_n}) \geq \F^2(\rho'_{PQ},\rho_{PQ})\geq 1-\eps^2$ and similarly $\F^2(\rho'_P,\rho_P)\geq 1-\eps^2$. The claim now follows by triangle inequality for purified distance (Fact \ref{fact:trianglepurified}).
\end{proof}

\begin{claim}
\label{gentlepovm}
Consider a pure quantum state $\ket{\rho}_{ORA} = \sum_i \sqrt{p_i}\ket{i}_O \ket{\rho^i}_{RA}$ and an isometry $\cA  = \sum_i P_i \otimes \ket{i}_{O'}$, such that $0 \preceq P_i \preceq \id_A, \sum_i P_i^2 = \id_A$. Define the state $\ket{\rho'}_{ORAO'}\defeq \sum_i \sqrt{p_i}\ket{i}_O\ket{\rho^i}_{RA}\ket{i}_{O'}$ and let $q_i \defeq \Tr(P^2_i \rho^i_A)$. Then it holds that
$$\F(\rho'_{ORAO'},\cA\rho_{ORA}\cA^{\dagger})\geq \sum_i p_iq_i.$$
\end{claim}
\begin{proof}
Consider the state
$$\cA\ket{\rho}_{ORA} = \sum_{i,j}\sqrt{p_i}\ket{i}_O (\id_R\otimes P_j)\ket{\rho^i}_{RA}\ket{j}_{O'}.$$
We compute 
\begin{eqnarray*}
\F(\rho'_{ORAO'},\cA\rho_{ORA}\cA^{\dagger}) &&=\bigg|\bigg(\sum_{i'} \sqrt{p_{i'}}\bra{i'}_O\bra{\rho^{i'}}_{RA}\bra{i'}_{O'}\bigg)\cdot \\ && \bigg(  \sum_{i,j}\sqrt{p_i}\ket{i}_O (\id_R\otimes P_j)\ket{\rho^i}_{RA}\ket{j}_{O'}\bigg)\bigg|\\ &&= | \sum_i p_i\bra{\rho^i}_{RA}(\id_R\otimes P_i)\ket{\rho^i}_{RA}|\\ &&= \sum_i p_i \Tr(P_i\rho^i_A)\\ &&\geq \sum_i p_i \Tr(P^2_i\rho^i_A),
\end{eqnarray*}
where the last inequality follows from the fact that $P_i^2 \preceq P_i$, which is implied by $P_i \preceq \id_A$. This completes the proof by definition of $q_i$.
\end{proof}

\subsection{Asymptotic and i.i.d. analysis}

In this section, we consider the problem of quantum state redistribution of the quantum state $\ketbra{\Phi}^{\otimes n}_{RABC}$, for $n$ large enough. For this, we use the following result, which clarifies the asymptotic and i.i.d. properties of the max-relative entropy and the hypothesis testing relative entropy. 

\begin{fact}[\cite{TomHay13, li2014}]
\label{dmaxequi}
Let $\eps\in (0,1)$ and $n$ be an integer. Let $\rho^{\otimes n}, \sigma^{\otimes n}$ be quantum states. Define $V(\rho\|\sigma) = \Tr(\rho(\log\rho - \log\sigma)^2) - (\relent{\rho}{\sigma})^2$ and $\Phi(x) = \int_{-\infty}^x \frac{e^{-x^2/2}}{\sqrt{2\pi}} dx$. It holds that
\begin{eqnarray*}
\dmaxeps{\rho^{\otimes n}}{\sigma^{\otimes n}}{\eps} &=& n\relent{\rho}{\sigma} + \sqrt{nV(\rho\|\sigma)} \Phi^{-1}(\eps) \\ &+& O(\log n) ,
\end{eqnarray*}
and 
\begin{eqnarray*}
\dheps{\rho^{\otimes n}}{\sigma^{\otimes n}}{\eps} &=& n\relent{\rho}{\sigma} + \sqrt{nV(\rho\|\sigma)} \Phi^{-1}(\eps) \\ &+& O(\log n) .
\end{eqnarray*}
\end{fact}

We have the following theorem, where we have used the shorthands $R^n \equiv R\otimes R\otimes \ldots R$ and similarly for $B^n, A^n,C^n$. 
\begin{theorem}[Asymptotic i.i.d. analysis]
\label{asymptoticstateredist}
Fix an $\eps\in (0,1/9)$. There exists an entanglement-assisted one-way protocol $\cP$, which takes as input $\ket{\Phi}^{\otimes n}_{RACB}$ shared between three parties Reference ($R^n$), Bob ($B^n$) and Alice ($A^nC^n$) and outputs a state $\Phi'_{R^nA^nC^nB^n}$ shared between Reference ($R^n$), Bob ($B^nC^n$) and Alice ($A^n$) such that $\Phi'_{R^nA^nC^nB^n} \in  \ball{9\eps}{\Phi^{\otimes n}_{RACB}}$ Let the number of qubits communicated by Alice to Bob in $\cP$ be $Q(n,\eps)$. Then
\begin{eqnarray*}
\lim_{n\rightarrow \infty} \frac{1}{n}Q(n,\eps) \leq \frac{1}{2} \condmutinf{R}{C}{B}_{\Phi}. 
\end{eqnarray*}
\end{theorem}

\begin{proof}
Setting $\Phi_C^{\otimes n}\leftarrow\sigma_C$ and $\eps\leftarrow \eps_2, \eps_1$ in Theorem \ref{newcompression}, $Q(n,\eps)$ is upper bounded by
\begin{eqnarray*}
&&\frac{1}{2}\bigg(\inf_{\stackrel{\Phi'_{R^nB^nC^n}}{\in \ball{\eps}{\Phi^{\otimes n}_{RBC}}}}\dmax{\Phi'_{R^nB^nC^n}}{\Phi'_{R^nB^n}\otimes \Phi^{\otimes n}_C} - \sup_{\Phi''_{B^nC^n}\in \ball{\eps}{\Phi^{\otimes n}_{BC}}}\dheps{\Phi''_{B^nC^n}}{\Phi''_{B^n}\otimes \Phi^{\otimes n}_C}{\eps}\bigg) \\ &&+ 2\log\left(\frac{1}{\eps}\right).
\end{eqnarray*}

Setting $\Phi''_{B^nC^n} = \Phi^{\otimes n}_{BC}$, this can be upper bounded by 
\begin{eqnarray*}
&&\frac{1}{2}\bigg(\inf_{\stackrel{\Phi'_{R^nB^nC^n}}{\in \ball{\eps}{\Phi^{\otimes n}_{RBC}}}}\dmax{\Phi'_{R^nB^nC^n}}{\Phi'_{R^nB^n}\otimes \Phi^{\otimes n}_C}  - \dheps{\Phi^{\otimes n}_{BC}}{\Phi^{\otimes n}_{B}\otimes \Phi^{\otimes n}_C}{\eps}\bigg) \\ &&+ 2\log\left(\frac{1}{\eps}\right).
\end{eqnarray*}

Let $\Phi^*_{R^nB^nC^n}\in \ball{\frac{\eps}{2}}{\Phi^{\otimes n}_{RBC}}$ be the quantum state achieving the minimum in the definition of $$\dmaxeps{\Phi^{\otimes n}_{RBC}}{\Phi^{\otimes n}_{RB}\otimes \Phi^{\otimes n}_C}{\frac{\eps}{2}}.$$ Using Fact \ref{dmaxequivalence} with $\frac{\eps}{2}\leftarrow \eps$, $\Phi^{\otimes n}_{RB}\leftarrow\sigma_A$, $\Phi^{\otimes n}_C\leftarrow\sigma_B$, $\Phi^*_{R^nB^nC^n} \leftarrow \psi_{AB}$, we can further upper bound $Q(n,\eps)$ by
\begin{eqnarray*}
&&\frac{1}{2}\bigg(\dmaxeps{\Phi^{\otimes n}_{RBC}}{\Phi^{\otimes n}_{RB}\otimes \Phi^{\otimes n}_C}{\frac{\eps}{2}} - \dheps{\Phi^{\otimes n}_{BC}}{\Phi^{\otimes n}_{B}\otimes \Phi^{\otimes n}_C}{\eps}\bigg) + O\left(\log\left(\frac{1}{\eps}\right)\right).
\end{eqnarray*} 
The theorem now follows by invoking Fact \ref{dmaxequi}. 
\end{proof}

\section{Connecting hypothesis testing relative entropy with fidelity}
\label{sec:dhdhalf}

In order to compare our bound with the existing results, we shall connect the hypothesis testing relative entropy with fidelity between quantum states. We prove the following theorem. 
\begin{theorem}
\label{theo:dhdhalf}
Let $\rho_1,\rho_2$ be quantum states and $\eps\in (0,1)$ be an error parameter. It holds that
$$\dheps{\rho_1}{\rho_2}{0}\leq \dhalf{\rho_1}{\rho_2} \leq \dheps{\rho_1}{\rho_2}{\eps} + \log\frac{4}{\eps}.$$
\end{theorem}

\begin{proof}
We first show the lower bound on $\dhalf{\rho_1}{\rho_2}$. Let $\Pi$ be an operator satisfying $\Tr(\Pi\rho_1)=1$ and $\Tr(\Pi\rho_2) = 2^{-\dheps{\rho_1}{\rho_2}{0}}$. Consider the measurement $\Lambda(X) = \Tr(\Pi X)\ketbra{0} + (1-\Tr(\Pi X))\ketbra{1}$. By monotonicity of fidelity under quantum operations (Fact \ref{fact:monotonequantumoperation}), 
\begin{eqnarray*}
\F(\rho_1,\rho_2) &\leq& \F(\Lambda(\rho_1), \Lambda(\rho_2)) = 2^{-\frac{1}{2}\dheps{\rho_1}{\rho_2}{0}}. 
\end{eqnarray*}
Thus, $\dheps{\rho_1}{\rho_2}{0} \leq -2\log\F(\rho_1,\rho_2)$, which leads to the desired lower bound.

To prove the upper bound, we proceed as follows. For a parameter $p$ to be chosen later, let $\rho\defeq p\rho_1 + (1-p)\rho_2$. Let $p_1\defeq p$ and $p_2\defeq 1-p$. Define the operators $\Lambda_1 \defeq \sqrt{p_1\rho^{-\frac{1}{2}}\rho_1\rho^{-\frac{1}{2}}}$ and $\Lambda_2 \defeq \sqrt{p_2\rho^{-\frac{1}{2}}\rho_2\rho^{-\frac{1}{2}}}$.  From Fact \ref{prettygoodmeas}, we have that 
\begin{eqnarray*}
\sum_{i=1}^2p_i\Tr(\Lambda_i^2\rho_i) &=& \sum_{i=1}^2p_i^2\Tr(\rho^{-\frac{1}{2}}\rho_i\rho^{-\frac{1}{2}}\rho_i) \geq 1- 2\sqrt{p_1p_2}\F(\rho_1,\rho_2) \\ &=& 1-2\sqrt{p(1-p)}\F(\rho_1,\rho_2).
\end{eqnarray*}
Now, using the relation $\Lambda_1^2+\Lambda_2^2 = \id$, we can rewrite above equation as
$$p\Tr(\Lambda_2^2\rho_1)+ (1-p)\Tr(\Lambda_1^2\rho_2) \leq 2\sqrt{p(1-p)}\F(\rho_1,\rho_2).$$
This implies that $$p\Tr(\Lambda_2^2\rho_1) \leq 2\sqrt{p(1-p)}\F(\rho_1,\rho_2)$$ and $$(1-p)\Tr(\Lambda_1^2\rho_2) \leq 2\sqrt{p(1-p)}\F(\rho_1,\rho_2).$$
Thus, we obtain
 $$\Tr(\Lambda_2^2\rho_1) \leq \sqrt{\frac{4(1-p)}{p}\F^2(\rho_1,\rho_2)}$$ and $$\Tr(\Lambda_1^2\rho_2) \leq \sqrt{\frac{4p}{1-p}\F^2(\rho_1,\rho_2)}.$$
Now, for the $\eps$ as given in the statement of the theorem we choose $p$ such that $$\frac{4(1-p)}{p}\F^2(\rho_1,\rho_2) = \eps^2.$$To see that this choice is possible for every $\eps \in (0,1)$, we rewrite above equation as $$p = \frac{1}{1+ \frac{\eps^2}{4\cdot F^2(\rho_1,\rho_2)}}.$$
Then we obtain $\Tr(\Lambda_1^2\rho_1) = 1-\Tr(\Lambda_2^2\rho_1) = 1-\eps$ and $$\Tr(\Lambda_1^2\rho_2) \leq \frac{4\F^2(\rho_1,\rho_2)}{\eps} = 2^{-\dhalf{\rho_1}{\rho_2} + \log\frac{4}{\eps}}.$$ The lemma concludes by definition of $\dheps{\rho_1}{\rho_2}{\eps}$.

\end{proof}

An immediate corollary of Theorem \ref{theo:dhdhalf} is the following.
\begin{cor}
\label{bipartitedhdhalf} Let $\Phi_{BC}, \sigma_C$ be quantum states and $\eps\in (0,1)$. Then 
\begin{eqnarray*}
&&\sup_{\Phi'_{BC}\in \ball{\eps}{\Phi_{BC}}} \dhalf{\Phi'_{BC}}{\Phi'_B\otimes \sigma_C} \leq \sup_{\Phi'_{BC}\in \ball{\eps}{\Phi_{BC}}} \dheps{\Phi'_{BC}}{\Phi'_B\otimes \sigma_C}{\eps} + \log\frac{4}{\eps}.
\end{eqnarray*}
\end{cor}

\section{Comparison with previous work}
\label{sec:compare}

\subsection{Comparision of the achievability bounds}
\label{subsec:compare}

In \cite{Berta14}, following achievability bound was shown for quantum state redistribution of $\Phi_{RABC}$ with error $5\sqrt{\eps_1}+2\sqrt{\eps_2}$, for $\eps_1,\eps_2 \in (0,1)$ satisfying $5\sqrt{\eps_1}+2\sqrt{\eps_2}\leq 1$:

$$\frac{1}{2}\left(\hmaxeps{C}{B}{\eps_1}_{\Phi} - \hmineps{C}{RB}{\eps_2}_{\Phi}\right) + \mathcal{O}(\log\frac{1}{\eps_1\cdot\eps_2}).$$
In this section, we show that this quantity is an upper bound on our achievability result obtained in Theorem \ref{newcompression}. For this, setting $\sigma_C = \mu_C$ (where $\mu_C$ is the maximally mixed state on register $C$) in Theorem \ref{newcompression} and using Corollary \ref{bipartitedhdhalf}, we have the following upper bound on achievable quantum communication cost for quantum state redistribution with error $3\eps_1+6\eps_2$:
\begin{eqnarray*}
&&\frac{1}{2}\bigg(\inf_{\Phi'_{RBC}\in \ball{\eps_1}{\Phi_{RBC}}}\dmax{\Phi'_{RBC}}{\Phi'_{RB}\otimes \mu_C} - \sup_{\Phi''_{BC}\in \ball{\eps_2}{\Phi_{BC}}}\dhalf{\Phi''_{BC}}{\Phi''_{B}\otimes \mu_C}\bigg) \\ &&+ \log\frac{4}{\eps_1\cdot \eps^2_2}.
\end{eqnarray*}
Following is the main result of this section.
\begin{theorem}
\label{thm:compare}
Let $\eps_1,\eps_2 \in (0,1)$. It holds that
\begin{eqnarray*}
&&\inf_{\Phi'_{RBC}\in \ball{\eps_1}{\Phi_{RBC}}}\dmax{\Phi'_{RBC}}{\Phi'_{RB}\otimes \mu_C}  - \sup_{\Phi''_{BC}\in \ball{\eps_2}{\Phi_{BC}}}\dhalf{\Phi''_{BC}}{\Phi''_{B}\otimes \mu_C} \\ &&\leq \hmaxeps{C}{B}{\eps_2}_{\Phi}  -\hmineps{C}{RB}{\eps_1/2}_{\Phi}+ \log\frac{12}{\eps^2_1}.
\end{eqnarray*}
\end{theorem}
\begin{proof}
Using Fact \ref{dmaxequivalence}, we have that 
\begin{eqnarray*}
&&\inf_{\Phi'\in \ball{\eps_1}{\Phi}}\dmax{\Phi'_{RBC}}{\Phi'_{RB}\otimes \mu_C} \leq \inf_{\Phi'\in \ball{\eps_1/2}{\Phi}}\inf_{\sigma_{RB}}\dmax{\Phi'_{RBC}}{\sigma_{RB}\otimes \mu_C} + \log\frac{12}{\eps^2_1}.
\end{eqnarray*}
But, the definition of conditional min-entropy implies that
\begin{eqnarray*}
 \inf_{\sigma_{RB}}\dmax{\Phi'_{RBC}}{\sigma_{RB}\otimes \mu_C} &=& \inf_{\sigma_{RB}}\dmax{\Phi'_{RBC}}{\sigma_{RB}\otimes \id_C} + \log d_C \\ &=& -\hmin{C}{RB}_{\Phi'} + \log d_C.
\end{eqnarray*}
Thus, 
\begin{eqnarray}
\label{eq:imaxupperbound}
\inf_{\Phi'\in \ball{\eps_1}{\Phi}}\dmax{\Phi'_{RBC}}{\Phi'_{RB}\otimes \mu_C} &\leq& \inf_{\Phi'\in \ball{\eps_1/2}{\Phi}}
-\hmin{C}{RB}_{\Phi} + \log d_C + \log\frac{12}{\eps_1^2} \nonumber\\ &=& -\sup_{\Phi'\in \ball{\eps_1/2}{\Phi}}
\hmin{C}{RB}_{\Phi} + \log d_C + \log\frac{12}{\eps_1^2}\nonumber \\ &=& -\hmineps{C}{RB}{\eps_1/2}_{\Phi}+ \log d_C + \log\frac{12}{\eps^2_1}.
\end{eqnarray}
On the other hand,
\begin{eqnarray*}
&&- \dhalf{\Phi''_{BC}}{\Phi''_{B}\otimes \mu_C} = \log\F^2(\Phi''_{BC},\Phi''_B\otimes \mu_C) = \log\F^2(\Phi''_{BC},\Phi''_B\otimes \id_C) - \log d_C.
\end{eqnarray*}
But, from the definition of conditional max-entropy, $\log\F^2(\Phi''_{BC},\Phi''_B\otimes \id_C) \leq \hmax{C}{B}_{\Phi''}$. Thus,
\begin{eqnarray*}
- \sup_{\Phi''\in \ball{\eps_2}{\Phi}}\dhalf{\Phi''_{BC}}{\Phi''_{B}\otimes \mu_C}  &\leq& \min_{\Phi''\in\ball{\eps_2}{\Phi}}\hmax{C}{B}_{\Phi''} - \log d_C \\ &=& \hmaxeps{C}{B}{\eps_2}_{\Phi}-\log d_C. 
\end{eqnarray*}
Combining this with Equation \ref{eq:imaxupperbound}, we obtain 
\begin{eqnarray*}
&&\inf_{\Phi'\in \ball{\eps_1}{\Phi}}\dmax{\Phi'_{RBC}}{\Phi'_{RB}\otimes \mu_C} -  \sup_{\Phi''\in \ball{\eps_2}{\Phi}}\dhalf{\Phi''_{BC}}{\Phi''_{B}\otimes \mu_C} \\ &&\leq \hmaxeps{C}{B}{\eps_2}_{\Phi} - \hmineps{C}{RB}{\eps_1/2}_{\Phi}+ \log\frac{12}{\eps^2_1}.
\end{eqnarray*}
This completes the proof.
\end{proof}

Finally, we compare Theorem \ref{newcompression} to the main result of \cite{AnshuDJ14}, where the authors introduced the aforementioned technique of convex split and used it in the following result. Informally, it says that given a quantum state $\Phi_{RA'B'M}$ shared between Alice (registers $A'M$), Bob (register $B'$) and Reference (register $R$) during a quantum communication protocol, the message $M$ can be sent from Alice to Bob with communication cost close to $\imax(RB':M)_{\Phi}$.  

\begin{theorem}[\cite{AnshuDJ14}]
\label{oldcompression}
Fix $\eps\in (0,1)$. There exists an entanglement-assisted one-way protocol $\cP'$, which takes as input $\ket{\Phi}_{RA'MB'}$ shared between three parties Reference ($R$), Bob ($B'$) and Alice ($A'M$) and outputs a state $\Phi'_{RA'MB'}$ shared between Reference ($R$), Bob ($B'M$) and Alice ($A'$) such that $\Phi'_{RA'MB'} \in  \ball{\eps}{\Phi_{RA'MB'}}$ and the number of qubits communicated by Alice to Bob in $\cP'$ is upper bounded by:
$$\frac{1}{2}\imax(RB':M)_{\Phi} + \log\left(\frac{1}{\varepsilon}\right) .$$ 
\end{theorem}

Using above Theorem, it was shown that the following quantity tightly captures the quantum communication cost of quantum state redistribution (upto an additive factor of $\log \frac{1}{\eps}$) with error $\eps$:

\begin{definition}[\cite{AnshuDJ14}] Let $\varepsilon \in (0,1)$ and $\ket{\Phi}_{RABC}$ be a pure state. Define,
\begin{eqnarray*}  
&&\mathrm{Q}^{\varepsilon}_{\ket{\Phi}_{RABC}} \defeq \inf_{T, U_{BCT},\sigma'_T,\kappa_{RBCT}} \imax(RB:CT)_{\kappa}  
\end{eqnarray*}
with the condition that $U_{BCT}$ is a unitary on registers $BCT$, $\sigma_T\in \mathcal{D}(T)$, $\kappa_{RB}=\Phi_{RB}$ and
$$(\id_R \otimes U_{BCT}) \kappa_{RBCT}(\id_R \otimes U^{\dagger}_{BCT}) \in \ball{\eps}{\Phi_{RBC}\otimes\sigma'_T}.$$ 
\end{definition}

This quantity has the drawback of being a complex optimization problem, primarily because it is not clear what is an upper bound on the dimension of register $T$ and what is the nature of the unitaries $U_{BCT}$. Theorem \ref{newcompression} provides an explicit example of this unitary and ancillary register $T$, while retaining the techniques used in the achievability result of \cite{AnshuDJ14}.  

\subsection{Comparision of the shared entanglement} 

Quatum state splitting (in which register $B$ is trivial) has found important application in the question of Quantum Reverse Shannon Theorem \cite{BDHSW14}, as shown in \cite{Renner11}. Both these references noted the importance of using embezzling quantum states \cite{DamHayden03}, owing to the phenomena of entanglement spread \cite{HaydenWinter03, Harrow12} (for a quantum state $\ket{\psi}_{AB}$, its entanglement spread is $ \Delta(\psi_A):= H_0(\psi_A) - H_\infty(\psi_A)$, where $H_0(\psi_A)\defeq \log \Tr(\psi_A^0)$ and $H_\infty(\psi_A):= \log \frac{1}{\lambda_{\max}(\psi_A)}$, $\lambda_{\max}$ being the largest eigenvalue of $\psi_A$). To see why quantum state splitting of quantum state $\Phi_{RAC}$, with near optimal classical communication $\approx \dmax{\Phi_{RC}}{\Phi_R\otimes \Phi_C}$ \footnote{In this subsection, we will ignore the error parameter in all our discussions in order to keep the argument simple.}, is not possible using maximally entangled shared resources, we consider the change in entanglement spread on Bob's registers. Initially the entanglement spread is zero, as the marginal of shared entanglement on Bob's registers is maximally mixed. At the end of the protocol , it must be at least $\Delta(\Phi_C)$, which can be much larger than $2\cdot \dmax{\Phi_{RC}}{\Phi_R\otimes \Phi_C}$. This contradicts \cite[Theorem 1]{HaydenWinter03}, which states that the classical communication cost of a protocol is at least $\frac{1}{2}$ times the change in entanglement spread.

Above argument roughly explains the structure of shared entanglement in protocols constructed in the works \cite{BDHSW14, Renner11}. This is very different from the protocol constructed in Theorem \ref{newcompression} for quantum state splitting (or the achievability result in \cite{AnshuDJ14} for quantum state splitting), where the entanglement used is arguably simpler: many independent copies of the purification of quantum state $\Phi_C$. We briefly show how the structure of entanglement in our protocol fits with the discussion in previous paragraph, arguing that the solution lies in the fact that the quantum state $\Phi_C$ has sufficient entanglement spread. 

Ignoring the errors, consider the protocol in Theorem \ref{newcompression} for pure state $\ket{\Phi}_{RAC}$, with the purification $\ket{\Phi'}_{LC}$ of $\Phi_C$ serving as shared entanglement (and $L$ being the purifying register). We have the following transformation of the global quantum state. The initial quantum state between Alice ($ACL$), Bob $(BC')$ and Reference ($R$) is $\ket{\Phi}_{RAC}\otimes \ket{\Phi'}_{LC'}^{\otimes N}$ (for some $N$ and $C'\equiv C$). The final quantum state between Alice ($AL$), Bob $(BCC')$ and Reference ($R$) is $\ket{\Phi}_{RAC}\otimes \ket{\Phi'}_{LC'}^{\otimes N-1} \otimes \ket{\mu}_{JJ'}$, where $\mu$ is the maximally entangled quantum state. Since communication occurs between Alice and Bob, we consider the change in entanglement spread by considering two cases: $1)$ the reduced density matrix with Bob and $2)$ the reduced density matrix with Alice. 

In the first case, the initial entanglement spread is $N\Delta(\Phi_{C})$ and the final entanglement spread is $(N-1)\Delta(\Phi_{C}) + \Delta(\Phi_C)$. The change in entanglement spread is zero. The classical communication cost of the protocol is $\approx \dmax{\Phi_{RC}}{\Phi_R\otimes \Phi_C}$, which is positive and hence lower bounded by the change in entanglement spread. This is consistent with the main result in \cite[Theorem 1]{HaydenWinter03}. 

In the second case, the change in entanglement spread is $\Delta(\Phi_A)-\Delta(\Phi_{AC}) - \Delta(\Phi_C) = \Delta(\Phi_{RC}) - \Delta(\Phi_R) - \Delta(\Phi_C)$. Since $\supp(\Phi_{RC}) \subseteq \supp(\Phi_R\otimes \Phi_C)$, we obtain that the change in entanglement spread is at most $H_{\infty}(\Phi_C) + H_{\infty}(\Phi_R) - H_{\infty}(\Phi_{RC})$. Let $k_1\defeq \dmax{\Phi_{RC}}{\Phi_R\otimes \Phi_C}, k_2\defeq H_{\infty}(\Phi_C), k_3\defeq H_{\infty}(\Phi_R)$ and $k_4\defeq H_{\infty}(\Phi_{RC})$. Consider $$\Phi_{RC}\leq 2^{k_1}\Phi_R\otimes \Phi_C \leq 2^{k_1-k_2-k_3}\id_R\otimes \id_C,$$ by definition of $k_1,k_2,k_3$. This implies $2^{-k_4} \leq  2^{k_1-k_2-k_3}$, from which we conclude that $k_2+k_3- k_4 \leq k_1$.  This establishes the consistency with the main result in \cite[Theorem 1]{HaydenWinter03}.

\section{Conclusion}
We have presented a new achievability result for the task of entanglement-assisted quantum state redistribution, using the recently introduced techniques of convex-split \cite{AnshuDJ14} and position-based decoding \cite{AnshuJW17a}.  We have made comparison to the known result of Berta, Christandl and Touchette \cite{Berta14} and presented some new relations between quantum hypothesis testing divergence and sandwiched quantum R\'{e}nyi divergence of order $\frac{1}{2}$ in order to facilitate the comparison. 

An important question that we have not addressed in this work is the question of optimality of our protocol. Several lower bounds on the quantum communication cost of entanglement-assisted quantum state redistribution have been presented in \cite[Proposition 1]{Berta14} and \cite{LeditzkyWD16}, and it is not clear if they match with our achievability result. Further investigation may be needed to near-optimally capture the quantum communication cost of entanglement-assisted quantum state redistribution in the one-shot setting, in terms of an explicit or an easily characterized quantity (the near-optimal result in the reference \cite{AnshuDJ14} is not in terms of an explicit or an easily characterized quantity and it requires further understanding). Such a quantity could be viewed as a one-shot version of the conditional quantum mutual information, several candidates of which have been proposed in the work \cite{BertaSW15}.

\subsection*{Acknowledgment} 
We thank the anonymous referees for very helpful suggestions related to the manuscript.

This work is supported by the Singapore Ministry of Education and the National Research Foundation,
through the Tier 3 Grant “Random numbers from quantum processes” MOE2012-T3-1-009 and NRF RF Award NRF-NRFF2013-13. 

\bibliographystyle{alpha}
\bibliography{References}
\end{document}